\documentclass[conference]{IEEEtran}
\IEEEoverridecommandlockouts
\usepackage{cite}
\usepackage{amsmath,amssymb,amsfonts}
\usepackage{algorithmic}
\usepackage{graphicx}
\usepackage{textcomp}
\usepackage{xcolor}
\def\BibTeX{{\rm B\kern-.05em{\sc i\kern-.025em b}\kern-.08em
    T\kern-.1667em\lower.7ex\hbox{E}\kern-.125emX}}
\begin{document}

\title{Work in Progress: AI-Powered Engineering-Bridging Theory and Practice}

\author{
    \IEEEauthorblockN{Oz Levy}
    \IEEEauthorblockA{
        Faculty of Industrial Eng.\\
        and Technology Management - HIT \\
        Holon, Israel \\
        oz.levy@hit.ac.il
    }
    \and
    \IEEEauthorblockN{ Ilya Dikman}
    \IEEEauthorblockA{
         Faculty of Industrial Eng.\\
        and Technology Management - HIT \\
        Holon, Israel \\
        ilya.dikman@hit.ac.il
    }
    \and
    \IEEEauthorblockN{ Natan Levy}
    \IEEEauthorblockA{
        dept. Computer Science - HUJI \\
        Jerusalem, Israel \\
        Natan.Levy1@mail.huji.ac.il
    }
    \and

\IEEEauthorblockN{Michael Winokur}
    \hspace{20cm}
    \IEEEauthorblockA{
        Faculty of Industrial Eng.\\ 
        and Technology Management - HIT \\
        Holon, Israel \\
        michaelw@hit.ac.il
    }
}

\maketitle

\begin{abstract}
This paper explores how generative AI can help automate and improve key steps in systems engineering. It examines AI’s ability to analyze system requirements based on INCOSE’s “good requirement” criteria, identifying well-formed and poorly written requirements. The AI does not just classify requirements but also explains why some do not meet the standards. By comparing AI assessments with those of experienced engineers, the study evaluates the accuracy and reliability of AI in identifying quality issues. Additionally, it explores AI’s ability to classify functional and non-functional requirements and generate test specifications based on these classifications. Through both quantitative and qualitative analysis, the research aims to assess AI’s potential to streamline engineering processes and improve learning outcomes. It also highlights the challenges and limitations of AI, ensuring its safe and ethical use in professional and academic settings.
\end{abstract}

\begin{IEEEkeywords}
AI technology in Systems Engineering, Requirements Engineering, Quality Criteria for Requirements, Safe AI, AI enhanced test generation
\end{IEEEkeywords}

\section{Introduction}
In recent years, artificial intelligence (AI) has revolutionized various aspects of life and industry, becoming a key tool in processes once considered not applicable to automated mechanisms \cite{b10}. Leading organizations like Tesla and NASA have integrated AI into large-scale projects \cite{b1,b8,b9}. For instance, Baidya et al. \cite{b2} explored the opportunities and challenges of using AI in critical projects like robotics and aeronautics, where AI assists with data analysis and provides insights into complex engineering tasks. 
In systems engineering, AI is expected to enhance the ability to analyze and classify engineering requirements, which form the foundation for product design and development. Existing research has explored AI’s role in classifying requirements in software and quality management systems \cite{b17,b14}.
However, these studies primarily cover specific, categorized tasks in software or quality requirements rather than providing a tool for comprehensive, standardized classification as per INCOSE’s (\emph{International Council on Systems Engineering}) standards \cite{b7} and classify requirements into categories such as functional, non-functional.
With increasing demands for product quality and precision, analyzing requirements based on criteria like necessity, clarity, and verifiability is crucial.AI may enhance this process, offering rapid and accurate analyses that increase systems engineers' efficiency.  However, significant challenges remain due to AI’s tendency to generate misinterpretations, as seen in various applications \cite{b11,b12}. Such inaccuracies could have severe consequences in complex engineering projects \cite{b6,b13}. This study, therefore, proposes to evaluate the accuracy and reliability of AI models in such contexts, asking whether AI could match a systems engineer's judgment or remains merely an auxiliary tool, similar to calculators.
This research also examines the feasibility of integrating AI into engineering education, emphasizing how AI-based tools can equip engineers with 21st-century skills while fostering responsible engineering practices to address ethical challenges in critical systems. By bridging the gap between research and practical application, it aims to equip engineers with essential tools for today’s technological landscape.

\section{Methodology}
\begin{figure}[htbp]
    \centering
    \includegraphics[scale=5, trim=0cm 0cm 0cm 0cm, clip, width=75mm]{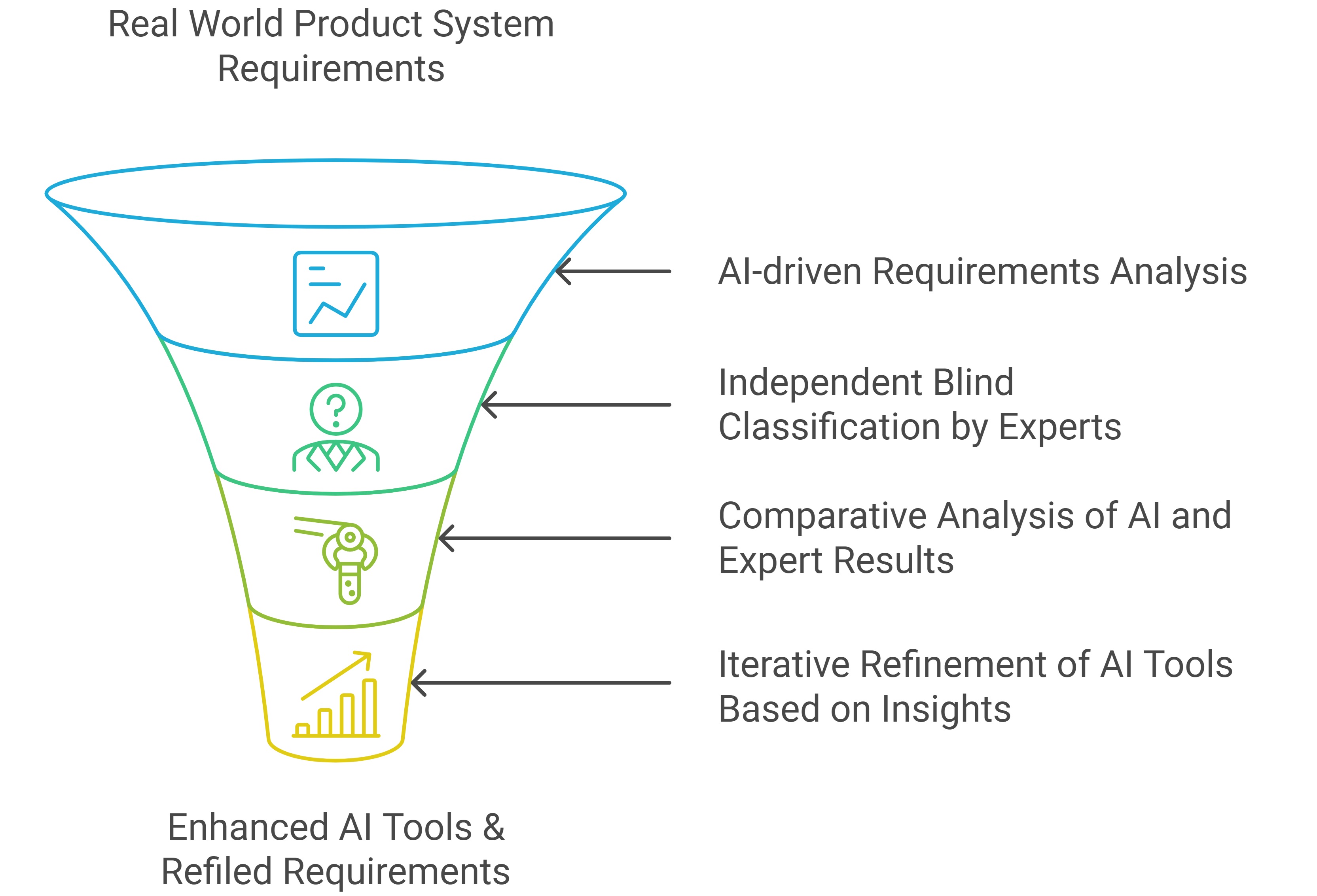}
    \caption{Logical Flow for AI-driven Requirements Analysis and Classification}
    \label{Figure1}
\end{figure}

According to the INCOSE \cite{b7} guide, a ''good requirement`` is defined by several key parameters that ensure clarity, precision, and feasibility.
The research investigates how the integration of Natural Language Processing (NLP) and Machine Learning (ML) techniques, such as text classification, clustering, and topic modeling using methods like Support Vector Machine (SVM), Convolutional Neural Networks (CNNs), and Latent Dirichlet Allocation (LDA), offers a way to automate requirements' analysis and classification.   
To classify system requirements as either functional or non-functional (quality requirements), An AI-driven approach will be adopted based on the Quality Requirements Mining and Classification Process methodology \cite{b14}. This method utilizes Natural Language Processing (NLP) techniques like Word2Vec and Doc2Vec, combined with Convolutional Neural Networks (CNNs). The AI will analyze the Software Requirements Specifications (SRS) and categorize requirements based on ISO/IEC 25030 standards, effectively distinguishing between functional and non-functional requirements, ensuring efficient processing of large SRS documents. Fig. \ref{Figure1} shows the logical flow proposed for AI- Driven Requirements Analysis and Classification
To utilize AI for creating automated test cases, the following methodology is recommended: Leveraging Natural Language Processing (NLP) to parse and analyze language in requirements documents, machine learning models—such as Long Short-Term Memory (LSTM) and Transformer-based models—extract key information to generate preliminary test cases \cite{b15}. This approach ensures that test cases comprehensively cover functional and non-functional requirements, aligning with industry standards like INCOSE’s criteria \cite{b16}.

\subsection{INCOSE's 'Good Requirement` Criteria}
The study focuses on seven out of nine parameters for a ''good requirement`` as defined by INCOSE:
\begin{enumerate}
    \item{Essential}: The requirement must be necessary.
    \item{Independent}: The requirement should express the need, not how to address it.
    \item{Unambiguous}: The requirement should have only one interpretation.
    \item{Complete}: The requirement must stand on its own.
    \item{Singular}: The requirement should express a single, clear idea.
    \item{Feasible}: The requirement must be implementable.
    \item{Verifiable}: The requirement must be verifiable.
\end{enumerate}

\subsection{Research Design}
The research aims to evaluate the capability of AI systems in classifying engineering requirements based on INCOSE's "good requirement" criteria. It also seeks to compare AI performance with expert classifications and iteratively improve the AI models.

The experiment will be executed through the following steps:
\begin{enumerate}
\item {Input- System Requirements Dataset:}
The experiment begins with a real-world product system requirements dataset sourced from the DR TOOL project. This dataset includes 31 stakeholder requirements and 76 optimized system requirements, representing critical operational functions (e.g., tracking mechanisms and alert systems) and non-functional attributes (e.g., reliability and security). Using real-world data ensures practical relevance and complexity.
\item {AI-Driven Requirements Analysis:}
AI models, specifically three state-of-the-art large language models (LLMs) - OpenAI's GPT-4\cite{b3}, Anthropic's Claude Sonnet\cite{b4}, and Meta's Llama\cite{b5} - will analyze the requirements dataset. These models will employ Natural Language Processing (NLP) and Machine Learning (ML) techniques to automatically classify requirements into functional and non-functional categories. Additionally, they will evaluate requirement quality according to INCOSE's established "good requirement" criteria, providing a comprehensive automated assessment framework.
\item {Independent Blind Classification by Experts:}
A panel of experienced systems engineers will independently classify the same dataset. This blind classification ensures unbiased human evaluation, serving as a benchmark for comparing AI-generated results.

\item {Comparative Analysis:}
The experiment will compare AI classifications with expert classifications using statistical measures, such as Cohen’s Kappa, to evaluate agreement. Discrepancies will be analyzed to identify strengths and limitations in the AI models.
\item {Iterative Refinement of AI Tools:}
Based on the comparative analysis, the AI models will be refined to address identified gaps. The improvements aim to enhance AI accuracy and utility in supporting systems engineers by identifying and analyzing problematic requirements.
 \end{enumerate}
\subsection{Application of AI Models in DR TOOL's Design}
\begin{figure}[htbp]
    \centering
    \includegraphics[scale=5, trim=0cm 0cm 0cm 0cm, clip, width=75mm]{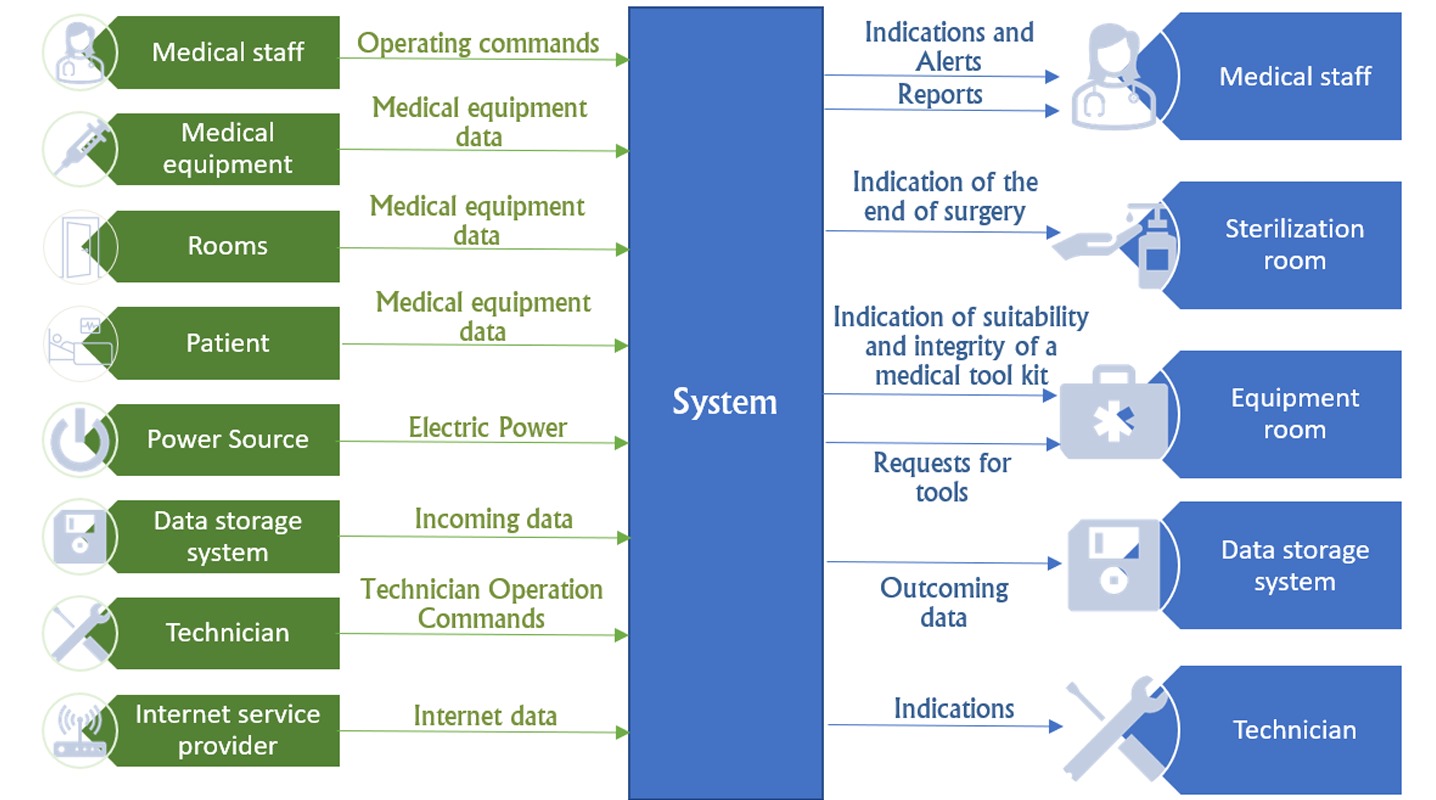}
    \caption{{The Dr. Tool System Diagram}}
    \label{Figure2}
\end{figure}

To assess the performance of AI models in classifying engineering requirements, The models were applied to the design project by DR TOOL \cite{b6}. This project focused on developing an RFID-based system for managing medical equipment in operating rooms, addressing key requirements for real-time tracking, equipment safety, and operational efficiency. The system was built on 31 stakeholder requirements and an additional 76 optimized system requirements, covering both functional and non-functional aspects of the system. The interaction of the Dr Tool System with its environment, including external players, inputs and outputs that drive the requirements is presented in its Context Diagram in Fig. \ref{Figure2} to illustrate the complexity of the test case.

By utilizing the real-world product system requirements dataset from DR TOOL’s system, The AI models’ ability to classify requirements is evaluated not only based on INCOSE’s "good requirement" criteria and by distinguishing between functional and non-functional requirements. This dataset includes comprehensive requirements, addressing critical operational functions such as tracking and alert mechanisms, alongside non-functional attributes like system reliability and security.
The evaluation not only enables a direct comparison between AI-generated classifications and those performed manually by experienced systems engineers but also provides insights into areas where AI models can be refined and improved. Furthermore, this process assists system engineers by identifying problematic requirements, offering justifications and actionable insights to enhance requirement quality. This dual benefit of improving AI tools and supporting engineers ensures a more efficient and effective requirements engineering process.

\subsection{Research Questions}
The research questions stem from the objective of the study. Key questions include:
\begin{enumerate}
    \item To what extent can AI classify engineering requirements according to INCOSE parameters of a ''good requirement``, compared to an experienced systems engineer?
    \item How effectively can AI distinguish between functional and non-functional requirements compared to an experienced systems engineer?
    \item What are the advantages and disadvantages of AI compared to a human expert in understanding and classifying engineering requirements?
\end{enumerate}

\section{Expected Contribution of the Research}
From a theoretical perspective, it is expected that AI models will demonstrate a certain level of proficiency in classifying engineering requirements according to INCOSE's criteria. However, they may struggle with 'hallucinations` and contextual misunderstandings that experienced engineers intuitively grasp. The research will contribute to a deeper understanding of the capabilities and limitations of AI models in analyzing and classifying engineering requirements, expanding knowledge in the field of systems engineering by integrating advanced AI technologies with traditional engineering processes.
From a practical perspective, the findings could lead to the development of new AI-based tools and work methods for systems engineers. Understanding AI’s capabilities in requirement classification may help increase efficiency, improve product quality, and reduce errors, budget overruns, and delays. Additionally, the findings may assist organizations in making informed decisions about incorporating AI into their engineering processes while accounting for the accompanying challenges and limitations.
\section{Discussion}
\subsection{Bridging Research and Practice}
The integration of AI tools in requirements analysis represents a significant step towards bridging the gap between theoretical research and practical application. By automating requirement classification and improving engineering workflows, this study demonstrates how AI can effectively transition from a research tool to a practical asset in industry. By evaluating the performance of AI models in real-world scenarios, the study provides insights into how AI can be effectively utilized in systems engineering processes.

\subsection{Education in the Age of Generative AI}
Integrating artificial intelligence into engineering education provides a unique opportunity to enhance students' understanding of high-quality requirements. By leveraging AI tools designed to classify requirements based on INCOSE standards, this research equips students with the ability to analyze "good" and "poor" requirements, supported by relevant justifications. This hands-on approach enables learners to identify gaps, refine their skills in crafting precise and effective requirements, and bridge the theoretical understanding of systems engineering with practical application. Furthermore, exposure to the strengths and limitations of AI models fosters critical thinking and ethical decision-making, essential for preparing engineers to responsibly integrate emerging technologies into their workflows.

\subsection{Responsible Engineering Education}
The research emphasizes responsible AI integration, ensuring that engineers are aware of the ethical considerations and potential risks associated with AI applications in critical systems. By fostering a culture of responsibility and ethical awareness, engineering education can produce professionals capable of making informed decisions.

\section{Future Work}
While current market-leading AI models have demonstrated promise across various applications, they are not specifically optimized for systems engineering tasks. To address this gap, future work could focus on adapting and training existing models for handling engineering requirements, thereby enhancing their effectiveness in this domain. This effort could involve refining algorithms and expanding datasets to better align with the distinct characteristics of engineering requirements. The development of a dedicated tool for requirements classification and analysis could be explored. Such a tool would provide structured guidelines and best practices for systems engineers, enabling more efficient and accurate handling of requirements. By pursuing these approaches, The aim is to enhance the integration of AI into systems engineering processes and improve overall project outcomes.
To translate the findings into industry practice, future work will focus on developing practical tools and training programs for systems engineers. This includes:
\begin{itemize}
    \item{Developing AI-Based Tools} Creating user-friendly AI tools that can assist engineers in requirement analysis while providing transparency in decision-making processes.
    \item{Training and Workshops} Organizing workshops and training sessions to educate engineers on the effective and responsible use of AI tools in their workflows.
    \item{Curriculum Integration} Incorporating AI and its applications into engineering curricula to prepare students with the necessary skills and knowledge.
\end{itemize}

\section{Conclusion}
This research underscores the AI potential in systems engineering while acknowledging its limitations. By combining AI tools with human expertise, it is possible to improve requirement analysis processes, enhance efficiency, and maintain lofty standards of quality and safety. The study contributes to bridging the gap between theory and practice, promoting responsible engineering education, and preparing engineers for the challenges of the modern technological era.\\

This research highlights the potential of AI to enhance systems engineering by automating requirements analysis and aligning with INCOSE standards. Future work will refine AI models for greater contextual understanding, apply them to more complex engineering tasks, and explore their integration into educational programs to further develop technical and ethical skills in engineers.

\vspace{12pt}

\end{document}